\documentclass[a4paper,12pt,oneside]{article}
\usepackage[sc]{mathpazo}
\usepackage[pdftex]{color,graphicx}
\usepackage[numbers,super,comma,sort&compress]{natbib}
\usepackage[tbtags]{amsmath}
\usepackage[bf,scriptsize]{caption}
\usepackage{pdfpages}
\usepackage{hyperref}

\DeclareGraphicsExtensions{.pdf,.png,.jpg}

\begin{document}

\title{Tackling Exascale Software Challenges in Molecular Dynamics Simulations\\ with GROMACS}

\author{Szil\'ard P\'all$^{1,2,\dagger}$ \and Mark James Abraham$^{1,2,\dagger}$ \and Carsten Kutzner$^{3}$ \and Berk Hess$^{1,2}$ \and Erik Lindahl$^{1,2,4}$}

\date{Feb 12, 2015}

\maketitle

\thispagestyle{empty}
\noindent
\small{
\vskip10em
$^1$Science for Life Laboratory, Stockholm and Uppsala, 171 21 Stockholm, Sweden, \\
$^2$Dept. Theoretical Physics, KTH Royal Institute of Technology, 10691 Stockholm, Sweden, \\
$^3$
Theoretical and Computational Biophysics Dept., 
Max Planck Institute for Biophysical Chemistry, Am Fassberg 11, 37077 G\"ottingen, Germany \\
$^4$Center for Biomembrane Research, Dept. Biochemistry \& Biophysics, Stockholm University, SE-10691 Stockholm, Sweden \\

\noindent
Address correspondence to: Erik Lindahl, \url{erik.lindahl@scilifelab.se}\\
}

\noindent
The final publication is available at\\
\url{http://link.springer.com/chapter/10.1007%2F978-3-319-15976-8_1}

\vskip5em

\noindent
$^{\dagger}$ The authors contributed equally.

\pagebreak

\section{Abstract}
GROMACS is a widely used package for biomolecular simulation, and over the last two decades it has evolved from small-scale efficiency to advanced heterogeneous acceleration and multi-level parallelism targeting some of the largest supercomputers in the world. Here, we describe some of the ways we have been able to realize this through the use of parallelization on all levels, combined with a constant focus on absolute performance. Release 4.6 of GROMACS uses SIMD acceleration on a wide range of architectures, GPU offloading acceleration, and both OpenMP and MPI parallelism within and between nodes, respectively. The recent work on acceleration made it necessary to revisit the fundamental algorithms of molecular simulation, including the concept of neighborsearching, and we discuss the present and future challenges we see for exascale simulation - in particular a very fine-grained task parallelism. We also discuss the software management, code peer review and continuous integration testing required for a project of this complexity.

\section{Introduction}

Molecular Dynamics simulation of biological macromolecules has evolved from a narrow statistical-mechanics method into a widely applied biophysical research tool that is used outside theoretical chemistry. Supercomputers are now as important as centrifuges or test tubes in chemistry. However, this success also considerably raises the bar for molecular simulation implementations - it is no longer sufficient to reproduce experimental results or e.g. show proof-of-concept relative scaling. To justify the substantial supercomputing resources required by many computational chemistry projects the most important focus today is simply absolute simulation performance and the scientific results achieved. Exascale computing has potential to take simulation to new heights, but the combination of challenges that face software preparing for deployment at the exascale to deliver these results are unique in the history of software. The days of simply buying new hardware with a faster clock rate and getting shorter times to solution with old software are gone. The days of running applications on a single core are gone. The days of heterogeneous processor design to suit floating-point computation are back again. The days of performance being bounded by the time taken for floating-point computations are ending fast. The need to design with multi-core and multi-node parallelization in mind at all points is here to stay, which also means Amdahl's law\cite{Amd1967} is more relevant than ever.\footnote{Amdahl's law gives a model for the expected (and maximum) speedup of a program when parallelized over multiple processors with respect to the serial version. It states that the achievable speedup is limited by the sequential part of the program.}

A particular challenge for biomolecular simulations is that the computational problem size is fixed by the geometric size of the protein and the atomic-scale resolution of the model physics. Most life science problems can be reduced to this size (or smaller). It is possible to simulate much larger systems, but it is typically not relevant. Second, the timescale of dynamics involving the entire system increases
much faster than the length scale, due to the requirement of sampling the exponentially larger number of ensemble microstates. This means that weak scaling is largely irrelevant for life science; to make use of increasing amounts of computational resources to simulate these systems, we have to rely either on strong-scaling software engineering techniques, or ensemble simulation techniques.

The fundamental algorithm of molecular dynamics assigns positions and velocities to every particle in the simulation system, and specifies the model physics that governs the interactions between particles. The forces can then be computed, which can be used to update the positions and velocities via Newton's second law, using a given finite time step. This numerical integration scheme is iterated a large number of times, and it generates a series of samples from the thermodynamic ensemble defined by the model physics. From these samples, observations can be made that confirm or predict experiment. Typical model physics have many components to describe the different kinds of bonded and non-bonded interactions that exist. The \emph{non-bonded interactions} between particles model behaviour like van der Waals forces, or Coulomb's law. The non-bonded interactions are the most expensive aspects of computing the forces, and the subject of a very large amount of research, computation and optimization.

\begin{figure}
\centerline{\includegraphics[width=0.58\textwidth]{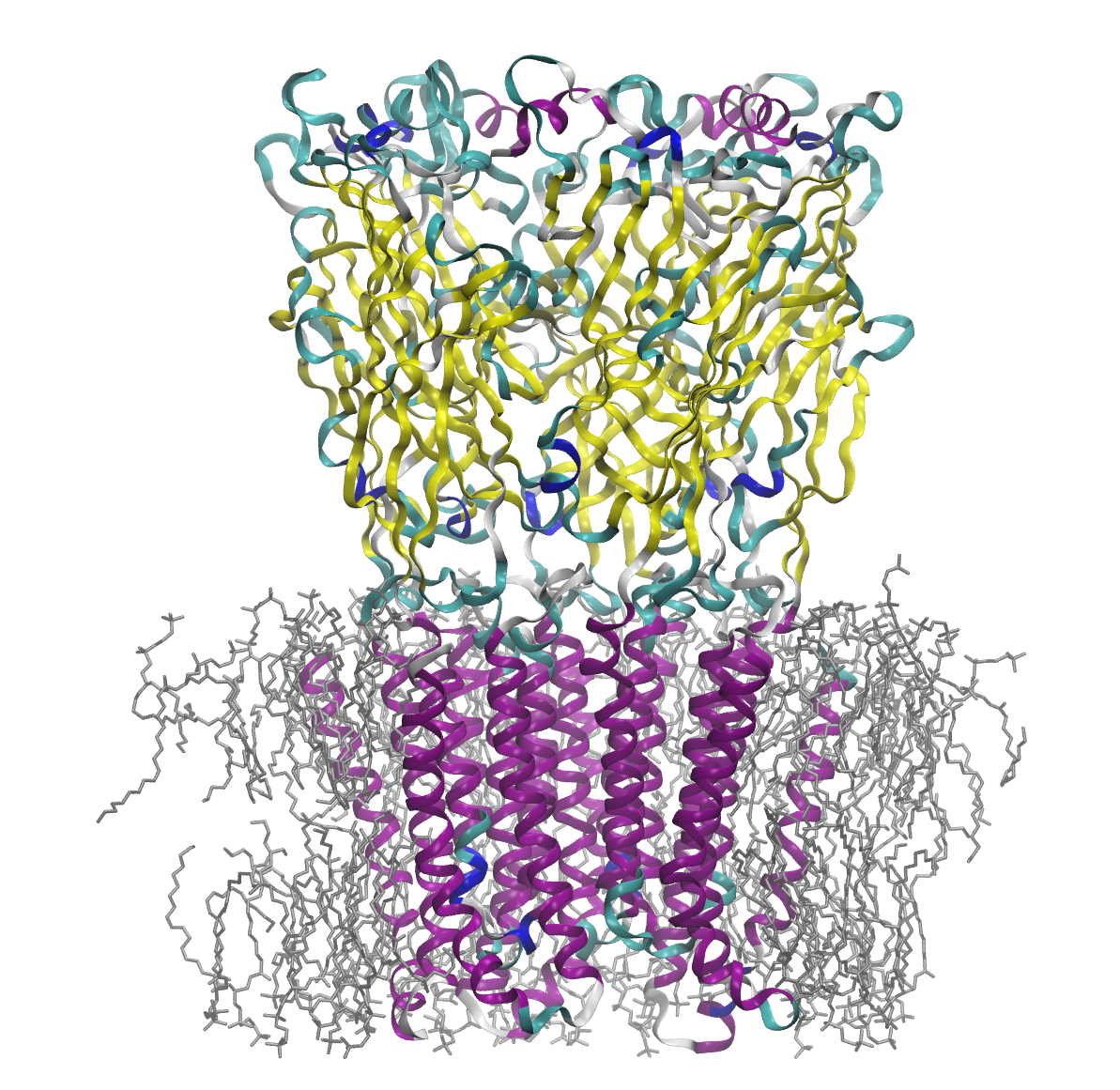}}
\caption{A typical GROMACS simulation system, featuring the ligand-gated ion-channel membrane protein GLIC (colored), embedded in a lipid membrane (grey). The whole system is solvated in water (not shown), giving a total of around 145,000 atoms. Image created with VMD.\cite{HumDalSch1996}}
\label{glic}
\end{figure}

Historically, the GROMACS molecular dynamics simulation suite has aimed at being a general-purpose tool for studying biomolecular systems, such as shown in Fig.~\ref{glic}. The development of the simulation engine focused heavily on maximizing single-core floating-point performance of its innermost compute kernels for non-bonded interactions. These kernels typically compute the electrostatic and van der Waals forces acting on each simulation particle from its interactions with all other inside a given spherical boundary. These kernels were first written in C, then FORTRAN, and later optimized in assembly language, mostly for commodity x86-family processors, because the data dependencies of the computations in the kernels were too challenging for C or FORTRAN compilers (then or now). The kernels were also specialized for interactions within and between water molecules, because of the prevalence of such interactions in biomolecular simulations. From one point-of-view, this extensive use of interaction-specific kernels can be seen as a software equivalent of application-specific integrated circuits.

Recognizing the need to build upon this good work by coupling multiple processors, GROMACS 4.0\cite{HesKutvan2008} introduced a minimal-communication neutral territory domain-decomposition (DD) algorithm,\cite{BowDroSha2005,BowDroSha2007} with fully dynamic load balancing. This spatial decomposition of the simulation volume created high-level data parallelism that was effective for near-linear scaling of the computation at around 400 atoms per core. The DD implementation required the use of MPI for message-passing parallel constructs. However, the needs of many simulation users can be met within a single node,\cite{ShiPan2000} and in that context the implementation overhead of MPI libraries was too high, not to mention it is difficult to employ in distributed computing. In GROMACS 4.5,\cite{ProPalSch2013} we implemented a multi-threaded MPI library with the necessary subset of the MPI API. The library has both POSIX and Windows threads back-ends (hence called thread-MPI) and uses highly efficient hardware-supported atomic and lock-free synchronization primitives.
This allows the existing DD implementation to work across multiple cores of a single node without depending on any external MPI library.

However, the fundamental limitation remained of a one-to-one mapping of MPI ranks to cores, and to domains. On the one hand, there is always a limit to how small a spatial domain can be, which will limit the number of domains the simulation box can be decomposed into, which in turn limits the number of cores that a parallelization with such a mapping can utilize. On the other hand, 
the one-to-one domains to cores mapping is cache-friendly as it creates independent data sets so that cores sharing caches can act without conflict, but the size of the volume of data that must be communicated so that neighboring domains act coherently grows rapidly with the number of domains. This approach is only scalable for a fixed problem size if the latency of communication between all cores is comparable and the communication book-keeping overhead grows only linearly with the number of cores. Neither is true, because network latencies are orders of magnitude higher than shared-cache latencies. This is clearly a major problem for designing for the exascale, where many cores, many nodes and non-uniform memory and communication latencies will be key attributes.

The other important aspect of the target simulations for designing for strong scaling is treating the long-range components of the atomic interactions. Many systems of interest are spatially heterogeneous on the nanometer scale (e.g. proteins embedded in membranes and solvated in water), and the simulation artefacts caused by failing to treat the long-range effects are well known. The \emph{de facto} standard for treating the long-range electrostatic interactions has become the “smooth particle-mesh Ewald” (PME) method,\cite{EssPerBer1995} whose cost for $N$ atoms scales as $N \log(N)$. A straightforward implementation where each rank of a parallel computation participates in an equivalent way leads to a 3D Fast Fourier Transform (FFT) that communicates globally. This communication quickly limits the strong scaling. To mitigate this, GROMACS 4.0 introduced a multiple-program multiple-data- (MPMD) implementation that dedicates some ranks to the FFT part; now only those ranks do all-to-all FFT communication. GROMACS 4.5 improved further by using a 2D pencil decomposition\cite{EleMorFit2003,Jag2005} in reciprocal space, within the same MPMD implementation. This coarse-grained task parallelism works well on machines with homogeneous hardware, but it is harder to port to accelerators or combine with RDMA constructs.

The transformation of GROMACS needed to perform well on exascale-level parallel hardware began after GROMACS 4.5. This requires radical algorithm changes, and better use of parallelization constructs from the ground up, not as an afterthought. More hands are required to steer the project, and yet the old functionality written before their time must generally be preserved. Computer architectures are evolving rapidly, and no single developer can know the details of all of them. In the following sections we describe how we are addressing some of these challenges, and our ongoing plans for addressing others.

\section{Handling exascale software challenges: algorithms and parallelization schemes}

\subsection{Multi-level parallelism}

Modern computer hardware is not only parallel, but exposes multiple levels of parallelism depending on the type and speed of data access and communication capabilities across different compute elements. For a modern superscalar CPU such as Intel Haswell, even a single core is equipped with 8 different execution ports, and it is not even possible to buy a single-core chip. Add hardware threads, complex communication crossbars, memory hierarchies, and caches larger than hard disks from the 1990s.
This results in a complex hierarchical organization of compute and communication/network elements from SIMD units and caches to network topologies, each level in the hierarchy requiring a different type of software parallelization for efficient use. HPC codes have traditionally focused on only two levels of parallelism: intra-node and inter-node. Such codes typically rely solely on MPI parallelization to target parallelism on multiple levels: both intra-socket, intra-node, and inter-node. This approach had obvious advantages before the multi-core and heterogeneous computing era when improvements came from CPU frequency scaling and evolution of interconnect. However, nowadays most scientific problems require complex parallel software architecture to be able use petaflop hardware efficiently and going toward exascale this is becoming a necessity. This is particularly true for molecular dynamics which requires reducing the wall-time per iteration to improve simulation performance.

On the lowest level, processors typically contain SIMD (single instruction multiple data) units which offer fine-grained data-parallelism through silicon dedicated to executing a limited set of instructions on multiple, currently typically 4-16, data elements simultaneously. Exploiting this low-level and  fine-grained parallelism has become crucial for achieving high performance, especially with new architectures like AVX and Intel MIC supporting wide SIMD. One level higher, multi-core CPUs have become the standard and several architectures support multiple hardware threads per core. Hence, typical multi-socket SMP machines come with dozens of cores capable of running 2-4 threads each (through simultaneous multi-threading, SMT, support). Simply running multiple processes (MPI ranks) on each core or hardware thread is typically less efficient than multi-threading. Achieving strong scaling in molecular dynamics requires efficient use of the cache hierarchy, which makes the picture even more complex. On the other hand, a chip cannot be considered a homogeneous cluster either. Accelerator coprocessors like GPUs or Intel MIC, often referred to as “many-core”, add another layer of complexity to the intra-node parallelism. These require fine-grained parallelism and carefully tuned data access patterns, 
as well as special programming models. Current accelerator architectures like GPUs also add another layer of interconnect in form of PCIe bus (Peripheral Component Interconnect Express) as well as a separate main memory. This means that data movement across the PCIe link often limits overall throughput. Integration of traditional latency-oriented CPU cores with throughput-oriented cores like those in GPUs or MIC accelerators is ongoing, but the cost of data movement between the different units will at least for the foreseeable future be a factor that needs to be optimized for.

Typical HPC hardware exhibits non-uniform memory access (NUMA) behavior on the node level: accessing data from different CPUs or cores of CPUs has a non-uniform cost. We started multithreading trials quite early with the idea of easily achieving load balancing, but the simultaneous introduction of NUMA suddenly meant a processor resembled a cluster internally.
Indiscriminately accessing memory across NUMA nodes will frequently lead to performance that is lower than for MPI.
Moreover, the NUMA behavior extends to other compute and communication components: the cost of communicating with an accelerator or through a network interface typically depends on the intra-node bus topology and requires special attention. On the top level, the interconnect links together compute nodes into a network topology. A side-effect of the multi-core evolution is that, while the network capacity (latency and bandwidth) per compute node has improved, the typical number of CPU cores they serve has increased faster; the capacity available per core has decreased substantially.

In order to exploit the capabilities of each level of hardware parallelism, a performance-oriented application needs to consider multiple levels of parallelism:
SIMD parallelism for maximizing single-core/thread performance;
multi-threading to exploit advantages of multi-core and SMT (e.g. fast data sharing);
inter-node communication-based parallelism (e.g. message passing with MPI); and
heterogeneous parallelism by utilizing both CPUs and accelerators like GPUs.

\begin{figure}
\centerline{\includegraphics[width=0.8\textwidth]{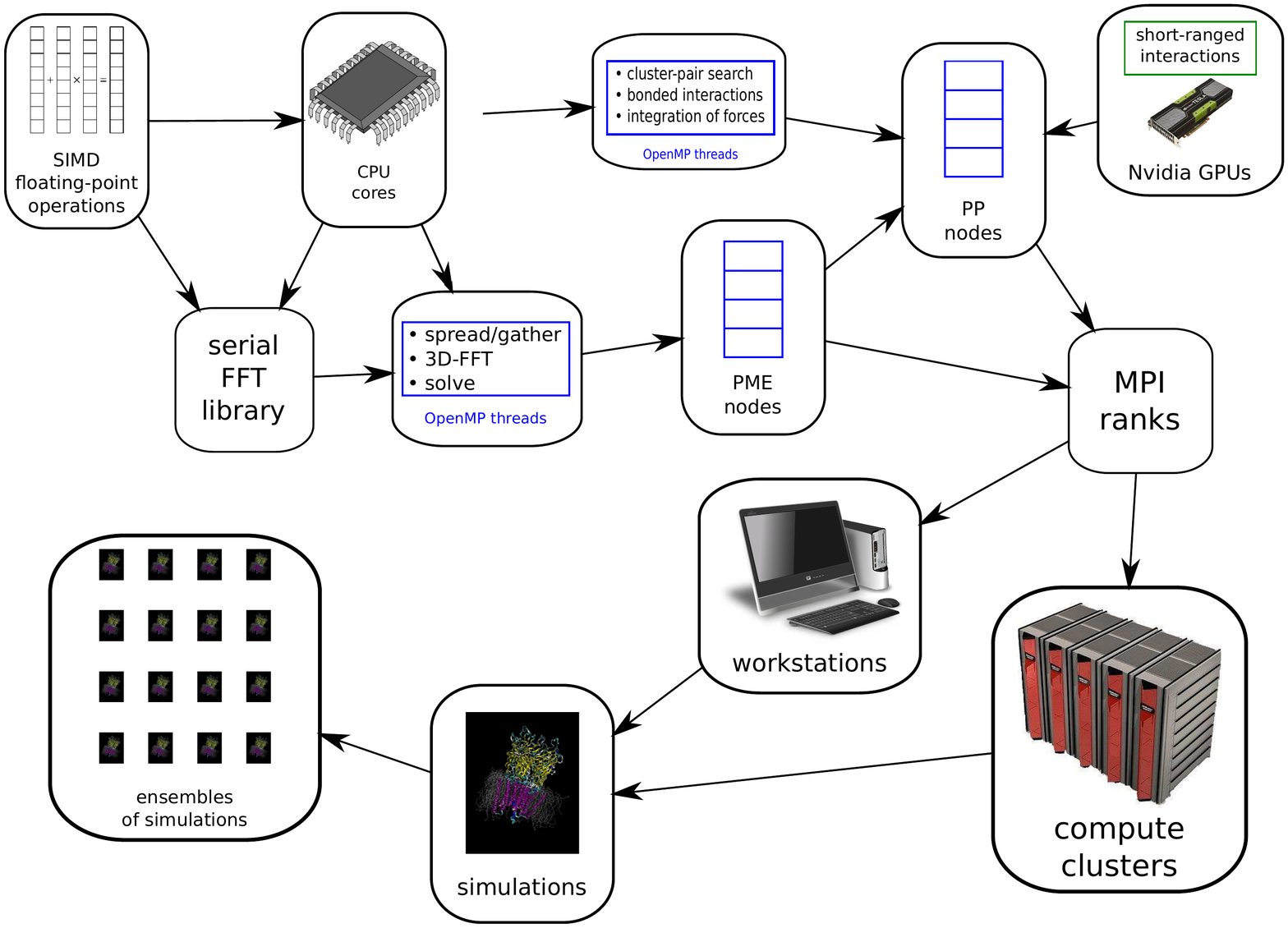}}
\caption{Illustration of multi-level parallelism in GROMACS 4.6. This exploits several kinds of fine-grained data parallelism, a multiple-program multiple-data (MPMD) decomposition separating the short-range particle-particle (PP) and long-range Particle Mesh Ewald (PME) force calculation algorithms,
coarse-grained data parallelism with domain-decomposition (DD) over  MPI ranks (implemented either on single-node workstations or compute clusters), and ensembles of related simulations scheduled e.g. by a distributed computing controller.\label{multilevelparallelism}}
\end{figure}

Driven by this evolution of hardware, we have initiated a re-redesign of the parallelization in GROMACS. In particular, recent efforts have focused on improvements targeting all levels of parallelization: new algorithms for wide SIMD and accelerator architectures, a portable and extensible SIMD parallelization framework, efficient multi-threading throughout the entire code, and an asynchronous offload-model for accelerators. The resulting multi-level parallelization scheme implemented in GROMACS 4.6 is illustrated in Fig.~\ref{multilevelparallelism}. In the following sections, we will give an overview of these improvements, highlighting the advances they provide in terms of making efficient use of current petascale hardware, as well as in paving the road towards exascale computing.

\subsection{SIMD parallelism}
All modern CPU and GPU architectures use SIMD-like instructions to achieve high flop rates. Any computational code that aims for high performance will have to make use of SIMD. For very regular work, such as matrix-vector multiplications, the compiler can generate good SIMD code, although manually tuned vendor libraries typically do even better.
But for irregular work, such as short-range particle-particle non-bonded interactions, the compiler usually fails since it cannot control data structures. If you think your compiler is really good at optimizing, it can be an eye-opening experience to look at the raw assembly instructions actually generated.  In GROMACS, this was reluctantly recognized a decade ago and SSE and Altivec SIMD kernels were written manually in assembly. These kernels were, and still are, extremely efficient for interactions involving water molecules, but other interactions do not parallelize well with SIMD using the standard approach of unrolling a particle-based Verlet-list.\cite{Verlet1967}

It is clear that a different approach is needed in order to use wide SIMD execution units like AVX or GPUs. We developed a novel approach, where particles are grouped into spatial clusters containing fixed number of particles.\cite{PalHes2013} First, the particles are placed on a grid in the $x$ and $y$ dimensions, and then binned in the $z$ dimension. This efficiently groups particles that are close in space, and permits the construction of a list of clusters, each containing exactly $M$ particles. A list is then constructed of all those cluster pairs containing particles that may be close enough to interact. This list of pairs of interacting clusters is reused over multiple successive evaluations of the non-bonded forces. The list is constructed with a buffer to prevent particle diffusion corrupting the implementation of the model physics.

The kernels that implement the computation of the interactions between two clusters $i$ and $j$ use SIMD load instructions to fill vector registers with copies of the positions of all $M$ particles in $i$. The loop over the $N$ particles in $j$ is unrolled according to the SIMD width of the CPU. Inside this loop, SIMD load instructions fill vector registers with positions of all $N$ particles from the $j$ cluster. This permits the computation of $N$ interactions between an $i$ and all $j$ particles simultaneously, and the computation of $M \times N$ interactions in the inner loop without needing to load particle data. With wide SIMD units it is efficient to process more than one $j$ cluster at a time.

$M$, $N$ and the number of $j$ clusters to process can be adjusted to suit the underlying characteristics of the hardware. Using $M$=1 and $N$=1 recovers the original Verlet-list algorithm. On CPUs, GROMACS uses $M$=4 and $N$=2, 4 or 8, depending on the SIMD width. On NVIDIA GPUs, we use $M$=8 and $N$=4 to calculate 32 interactions at once with 32 hardware threads executing in lock-step. To further improve the ratio of arithmetic to memory operations when using GPUs, we add another level of hierarchy by grouping 8 clusters together. Thus we store 64 particles in shared memory and calculate interactions with about half of these for every particle in the cluster-pair list.

The kernel implementations reach about 50\% of the peak flop rate on all supported hardware, which is very high for MD. This comes at the cost of calculating about twice as many interactions as required; not all particle pairs in all cluster pairs will be within the cut-off at each time step, so many interactions are computed that are known to produce a zero result. The extra zero interactions can actually be put to use as an effective additional pair list buffer additionally to the standard Verlet list buffer. As we have shown here, this scheme is flexible, since $N$ and $M$ can be adapted to current and future hardware. Most algorithms and optimization tricks that have been developed for particle-based pair lists can be reused for the cluster-pair list, although many will not improve the performance.

The current implementation of the cluster-based non-bonded algorithm already supports a wide range of SIMD instruction sets and accelerator architectures: 
SSE2, SSE4.1, AVX (256-bit and 128-bit with FMA), AVX2, BG/Q QPX, Intel MIC (LRBni), NVIDIA CUDA.
An implementation on a field-programmable gate array (FPGA) architecture is in progress. 

\subsubsection{Multi-threaded parallelism}\label{sec:multi-threading}
Before GROMACS 4.6, we relied mostly on MPI for both inter-node and intra-node parallelization over CPU cores. For MD this has worked well, since there is little data to communicate and at medium to high parallelization all data fits in L2 cache. Our initial plans were to only support OpenMP parallelization in the separate Particle Mesh Ewald (PME) MPI ranks. The reason for using OpenMP in PME was to reduce the number of MPI ranks involved in the costly collective communication for the FFT grid transpose. This 3D-FFT is the only part of the code that involves global data dependencies. Although this indeed greatly reduced the MPI communication cost, it also introduced significant overhead.

\begin{figure}[htp]
\centerline{\includegraphics[width=0.43\textwidth]{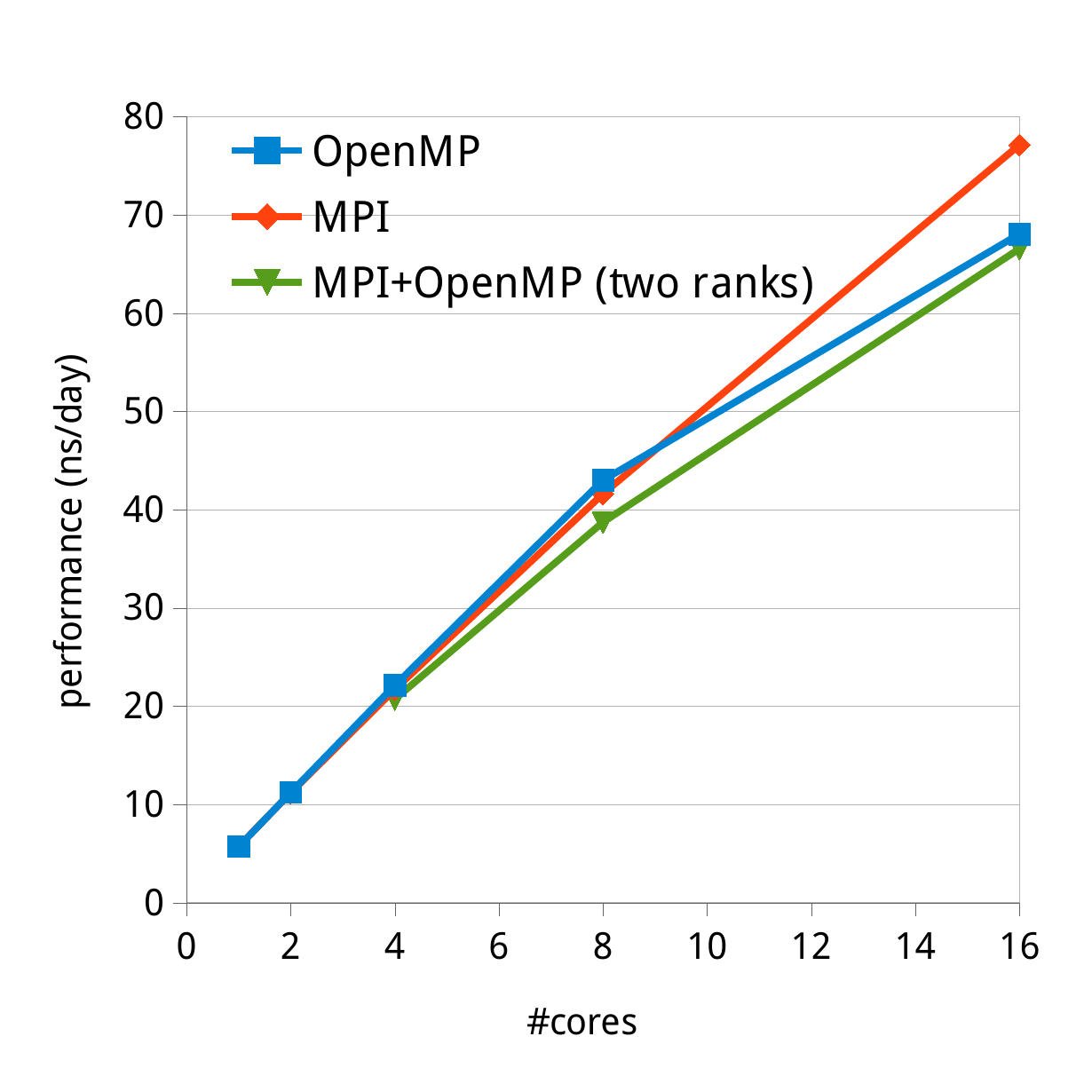}}
\caption{Comparison of single-node simulation performance using MPI, OpenMP, and combined MPI+OpenMP parallelization. The OpenMP multi-threading (blue) achieves the highest performance and near linear scaling up to 8 threads. It only deteriorates when threads on OpenMP regions need to communicate across the system bus. In contrast, the MPI-only runs (red) that require less communication scale well across sockets. Combining MPI and OpenMP parallelization with two ranks and varying the number of threads (green) results in worse performance due to the added overhead of the two parallelization schemes. Simulations were carried out on a dual-socket node with 8-core Intel Xeon E5-2690 (2.9 GHz Sandy Bridge). Input system: RNAse protein solvated in a rectangular box, 24k atoms, PME electrostatics, $0.9$ nm cut-off.
}\label{omp-tmpi}
\end{figure}

GROMACS 4.6 was designed to use OpenMP in all compute-intensive parts of the MD algorithm.\footnote{At the time of that decision, sharing a GPU among multiple MPI ranks was inefficient, so the only efficient way to use multiple cores in a node was with OpenMP within a rank. This constraint has since been relaxed.} Most of the algorithms are straightforward to parallelize using OpenMP. These scale very well, as Fig.~\ref{omp-tmpi} shows. Cache-intensive parts of the code like performing domain decomposition, or integrating the forces and velocities show slightly worse scaling.
Moreover, the scaling in these parts tends to deteriorate with increasing number of threads in an MPI rank -- especially with large number of threads in a rank, and when teams of OpenMP threads cross NUMA boundaries. 
When simulating at high ratios of cores/particles, each MD step can take as little as a microsecond. There are many OpenMP barriers used in the many code paths that are parallelized with OpenMP, each of which takes a few microseconds, which can be costly.

\begin{figure}[htp]
\centerline{\includegraphics[width=0.43\textwidth]{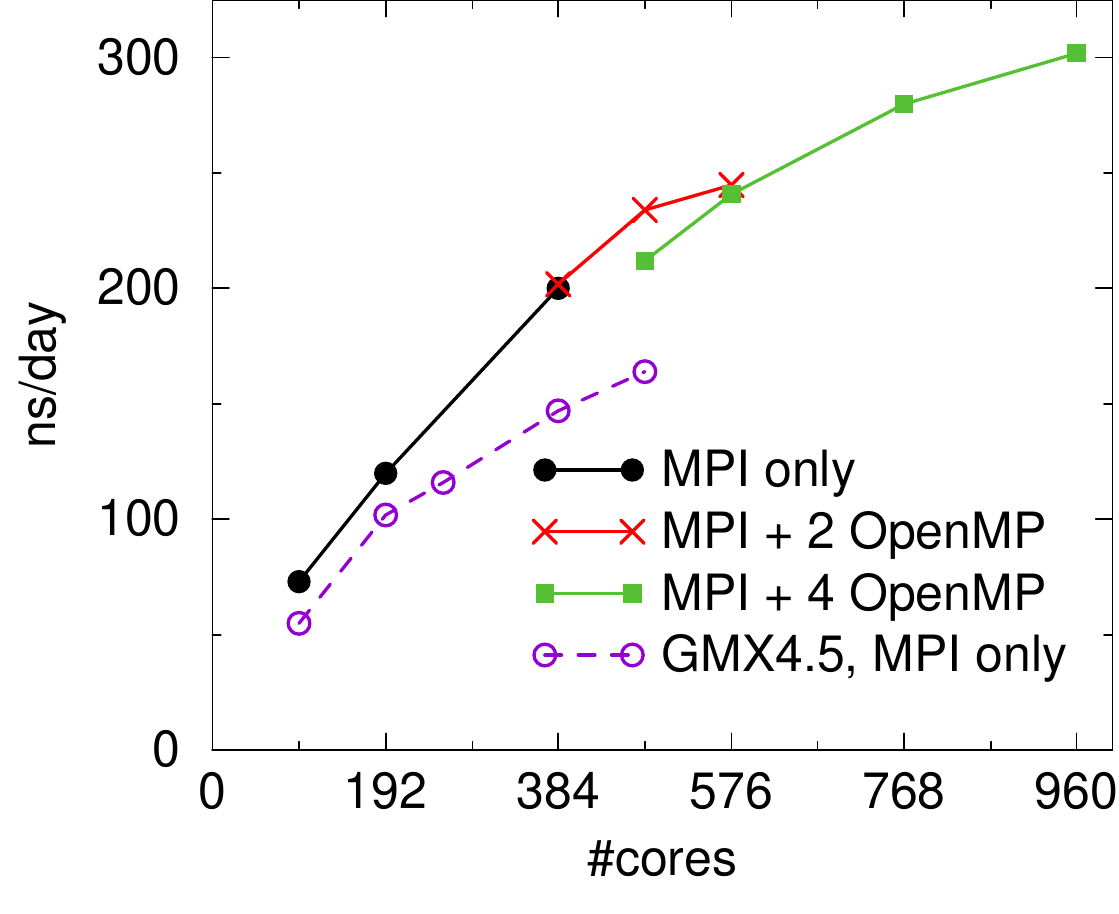}}
\caption{Improvements in strong scaling performance since GROMACS 4.5, using the $M \times N$ kernels and OpenMP parallelization in GROMACS 4.6. The plot shows simulation performance in ns/day for different software versions and parallelization schemes. Performance with one core per MPI rank is shown for GROMACS 4.5 (purple) and 4.6 (black). Performance with GROMACS 4.6 is shown using two (red) and four (green) cores per MPI rank using OpenMP threading within each MPI rank. Simulations were carried out on the Triolith cluster at NSC, using two 8-core Intel E5-2660 (2.2 GHz Sandy Bridge) processors per node and FDR Infiniband network. The test system is the GLIC membrane protein shown in Fig. 1 (144,000 atoms, PME electrostatics.)\label{glicscaling}}

\end{figure}

Accordingly, the hybrid MPI + OpenMP parallelization is often slower than an MPI-only scheme as Fig.~\ref{omp-tmpi} illustrates. Since PP (particle-particle) ranks only do low-volume local communication, the reduction in MPI communication from using the hybrid scheme is apparent only at high parallelization. There, MPI-only parallelization (e.g. as in GROMACS 4.5) puts a hard upper limit on the number of cores that can be used, due to algorithmic limits on the spatial domain size, or the need to communicate with more than one nearest neighbor. With the hybrid scheme, more cores can operate on the same spatial domain assigned to an MPI rank, and there is no longer a hard limit on the parallelization. Strong scaling curves now extend much further, with a more gradual loss of parallel efficiency. An example is given in Fig.~\ref{glicscaling}, which shows a membrane protein system scaling to twice as many cores with hybrid parallelization and reach double the peak performance of GROMACS 4.5. In some cases, OpenMP-only parallelization can be much faster than MPI-only parallelization if the load for each stage of the force computation can be balanced individually. A typical example is a solute in solvent, where the solute has bonded interactions but the solvent does not. With OpenMP, the bonded interactions can be distributed equally over all threads in a straightforward manner.

\subsubsection*{Heterogeneous parallelization}
Heterogeneous architectures combine multiple types of processing units, typically latency- and throughput-oriented cores -- most often CPUs and accelerators like GPUs, Intel MIC, or FPGAs. Many-core accelerator architectures have been become increasingly popular in technical and scientific computing mainly due to their impressive raw floating point performance. However, in order to efficiently utilize these architectures, a very high level of fine-grained parallelism is required. The massively parallel nature of accelerators, in particular GPUs, is both an advantage as well as a burden on the programmer. Since not all tasks are well suited for execution on the accelerators
this often leads to additional challenges for workload distribution and load balancing.
Moreover, current heterogeneous architectures typically use a slow PCIe bus to connect the hardware elements like CPUs and GPUs and move data between the separate global memory of each. This means that explicit data management is required. This adds a further latency overhead to challenge algorithms like MD that already face this as a parallelization bottle-neck.

GPU accelerators were first  supported experimentally in GROMACS with the OpenMM library,\cite{Eastman2010} which was
used as a black box to execute the entire simulation on the GPU. This meant that only a fraction of the diverse set of GROMACS algorithms were supported and simulations were limited to single-GPU use. Additionally, while OpenMM offered good performance for implicit-solvent models, the more common type of runs showed little speedup (and in some cases slowdown) over the fast performance on multi-core CPUs, thanks to the highly tuned SIMD assembly kernels.

\begin{figure}
\centerline{\includegraphics[width=0.5\textwidth]{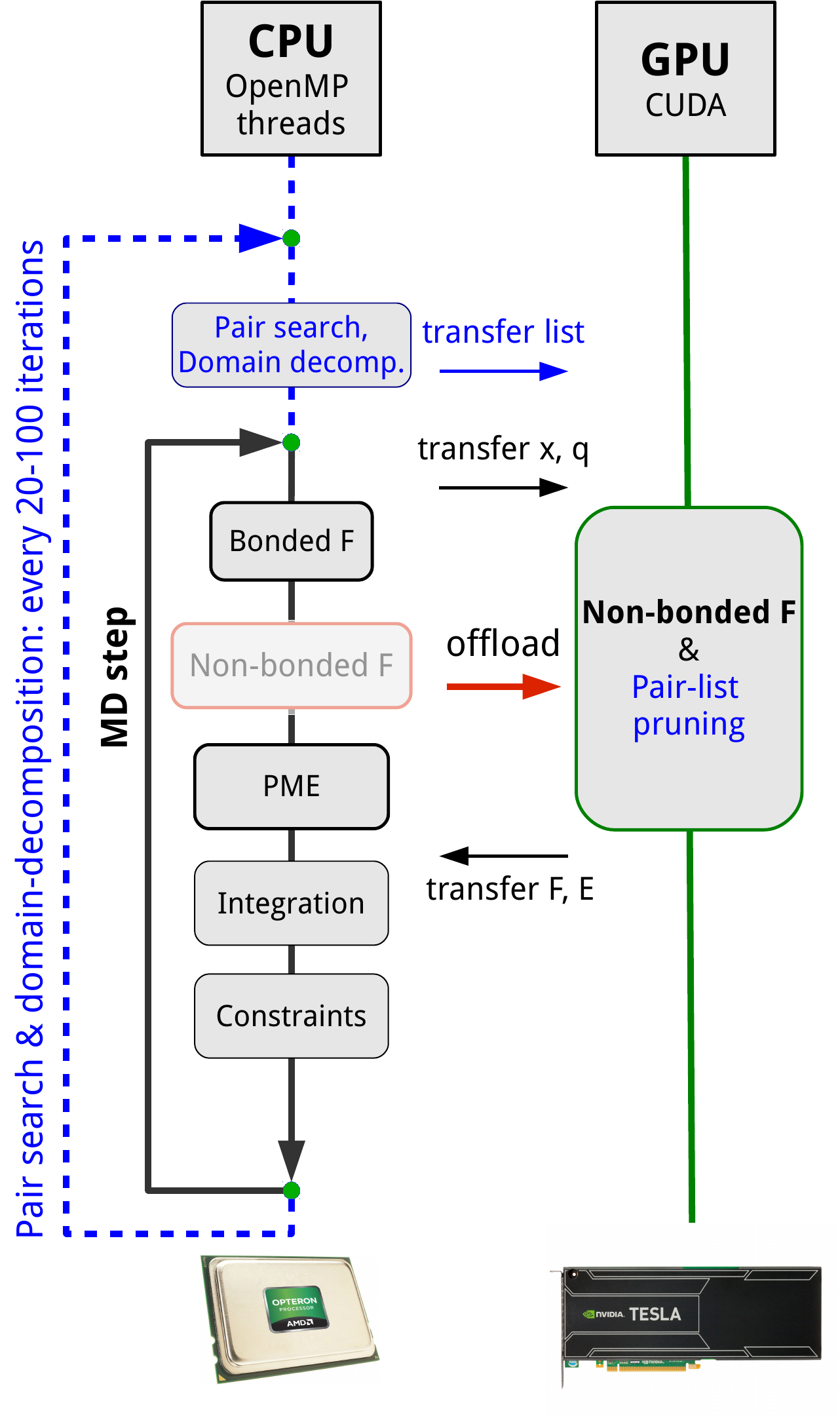}}
\caption{GROMACS heterogeneous parallelization using both CPU and GPU resources during each simulation time-step. The compute-heavy non-bonded interactions are offloaded to the GPU, while the CPU is responsible for domain-decomposition book-keeping, bonded force calculation, and lattice summation algorithms.
The diagram shows tasks carried out during a GPU-accelerated normal MD step (black arrows) as well as a step which includes the additional pair-search and domain-decomposition tasks are carried out (blue arrows). The latter, as shown above in blue, also includes an additional transfer, and the subsequent pruning of the pair list as part of the non-bonded kernel.\\
\tiny{ Source: \url{http://dx.doi.org/10.6084/m9.figshare.971161}. Reused under CC-BY; retrieved 22:15, March 23, 2014 (GMT).}
}
\label{gpu-acc}
\end{figure}

With this experience, we set out to provide native GPU support in GROMACS 4.6 with a few important design principles in mind. Building on the observation that highly optimized CPU code is hard to beat, our goal was to ensure that all compute resources available, both CPU and accelerators, are utilized to the greatest extent possible. We also wanted to ensure that our heterogeneous GPU acceleration supported most existing features of GROMACS in a single code base to avoid having to reimplement major parts of the code for GPU-only execution. This means that the most suitable parallelization is the offload model, which other MD codes have also employed successfully.\cite{Phillips2005,Brown2011} As Fig.~\ref{gpu-acc} illustrates, we aim to execute the compute-intensive short-range non-bonded force calculation on GPU accelerators, while the CPU computes bonded and long-range electrostatics forces, because the latter are communication intensive.

The newly designed future-proof SIMD-oriented algorithm for evaluating non-bonded interactions with accelerator architectures in mind has been discussed already.
It is highly efficient at expressing the fine-grained parallelism present in the pair-force calculation. Additionally, the atom cluster-based approach is designed for data reuse which is further emphasized by the super-cluster grouping. As a result, our CUDA implementation is characterized by a high ratio of arithmetic to memory operations which allows avoiding memory bottlenecks. These algorithmic design choices and the extensive performance tuning led to strongly instruction-latency bound CUDA non-bonded kernels, in contrast to most traditional particle-based GPU algorithms which are reported to be memory bound\cite{Anderson2008,Brown2011}.
Our CUDA GPU kernels also scale well, reaching peak pair-force throughput already around 20,000 particles per GPU.

In contrast to typical data-parallel programming for homogeneous CPU-only machines, heterogeneous
architectures require additional code to manage task scheduling and concurrent execution on the
different compute elements, CPU cores and GPUs in the present case. This is a complex component of our
heterogeneous parallelization which implements the data- and control-flow with the main goal of maximizing
the utilization of both CPU and GPU by ensuring optimal CPU-GPU execution overlap.

We combine a set of CPU cores running OpenMP threads with a GPU. As shown in Fig.~\ref{gpu-acc}, the
pair-lists required for the non-bonded computation are prepared on the CPU and transferred to the 
GPU where a pruning step is carried out after which the lists are reused for up to 100 iterations.
The extreme floating-point power of GPUs makes it feasible to use the much larger buffers required for this.
The transfer of coordinates, charges, and forces as well as compute kernels are
launched asynchronously as soon as data becomes available on the CPU. This ensures overlap of CPU and
GPU computation. Additional effort has gone into maximizing overlap by reducing the wall-time of
 CPU-side non-overlapping program parts through SIMD parallelization (in pair search and constraints)
 and efficient multi-threaded algorithms allowing GROMACS to achieve a typical CPU-GPU overlap of 60-80\%.

This scheme naturally extends to multiple GPUs by using the existing efficient neutral-territory do main-decomposition implemented using MPI
parallelization.
By default, we assign computation on each domain to a single GPU and a set of CPU cores. This typically means
 decomposing the system into as many domains as GPUs used, and running as many MPI ranks per node as GPUs in the node.
However, this will often require to run a large number of OpenMP threads in a rank (8-16 or even more with a single GPU per node),
potentially spanning across multiple NUMA domains. As explained in the previous section, this will lead to suboptimal multi-threaded
scaling -- especially affecting cache-intensive algorithms outside the CPU-GPU overlap region. To avoid this, multiple MPI ranks can
share a GPU, which reduces the number of OpenMP threads per rank.

\begin{figure}
\centerline{\includegraphics[width=0.5\textwidth]{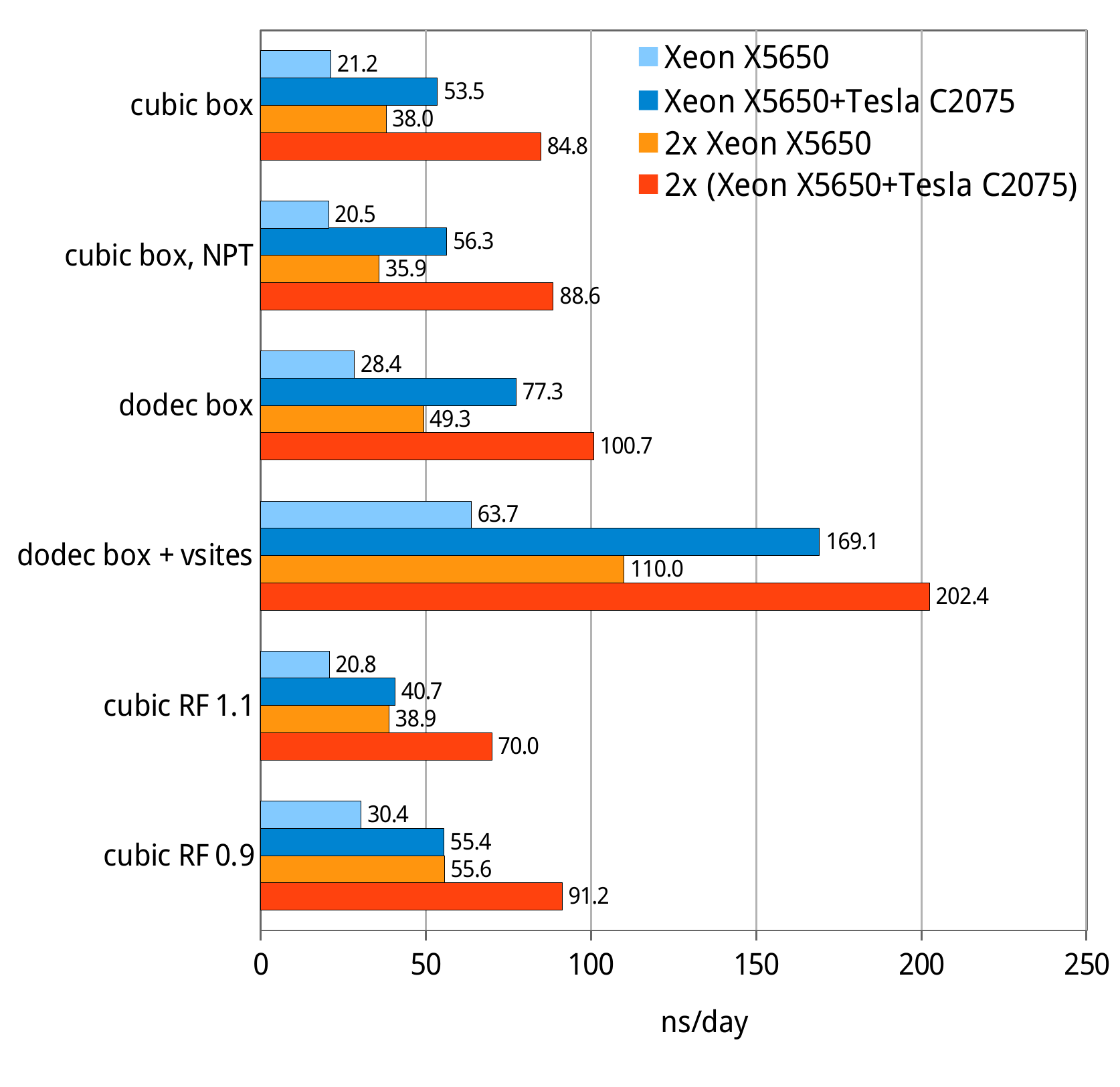}}
\caption{An important feature of the current heterogeneous GROMACS GPU implementation is that it works, and works efficiently, in combination with most other features of the software. GPU simulations can employ domain decomposition, non-standard boxes, pressure scaling, and virtual interaction sites to significantly improve the absolute simulation performance compared to the baseline.
Simulation system: RNAse protein solvated in rectangular (24 K atoms) and rhombic dodecahedron (16.8 k atoms) box, PME electrostatics, cut-off $0.9$ nm. Hardware: 2x Intel Xeon E5650 (2.67 GHz Westmere), 2x NVIDIA Tesla C2070 (Fermi) GPU accelerators.}
\label{rnase}
\end{figure}

\begin{figure}
\centerline{\includegraphics[width=0.5\textwidth]{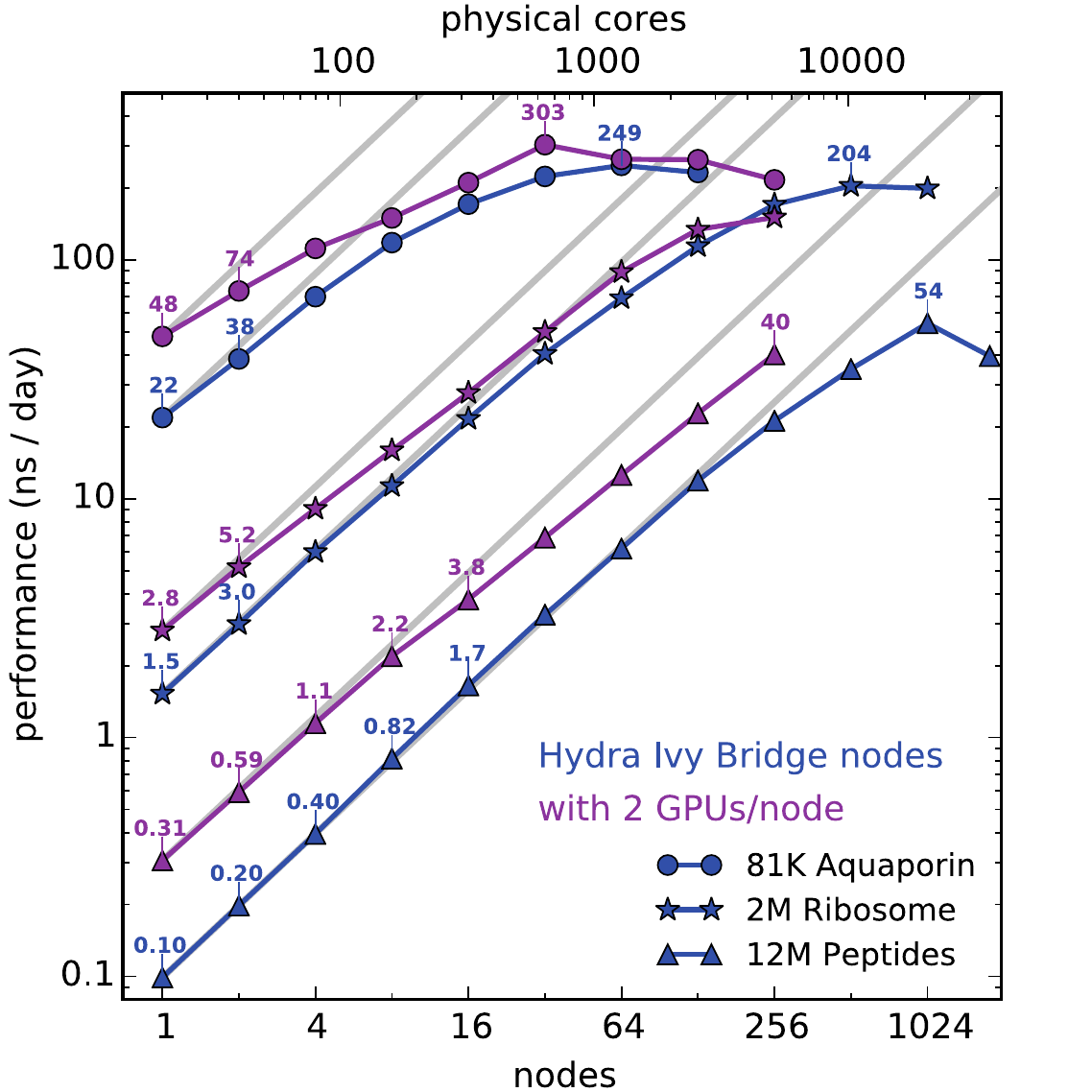}}
\caption{Strong scaling of GROMACS 4.6 on the HYDRA heterogeneous GPU-equipped machine in Garching, Germany. Grey lines indicate linear scaling. The hybrid version of GROMACS scales very well and achieves impressive absolute performance for both small and large systems. For the smaller systems, peak performance is achieved with 150 atoms per core, and the larger systems achieve sustained effective flop rate of 0.2 petaflops (only counting the number of useful floating-point operations, not the total). \\
Simulation systems: typical production systems of 81k atoms (circles), 2M atoms (stars), and 12M atoms (triangles) in size.
Hardware (per node): 2 10-core Xeon E5-2680v2 (2.8 GHz Ivy Bridge), 2 NVIDIA K20X, InfiniBand FDR14 (4 x 14 Gb/s) network.
}\label{garching}
\end{figure}

The heterogeneous acceleration in GROMACS delivers 3-4x speedup when comparing CPU only with CPU-GPU runs. Moreover, advanced features like arbitrary simulation box shapes and virtual interaction sites are all supported (Fig.~\ref{rnase}). Even though the overhead of managing an accelerator is non-negligible, GROMACS 4.6 shows great strong scaling in GPU accelerated runs reaching 126 atoms/core (1260 atoms/GPU) on common simulation systems (Fig.~\ref{garching}).

Based on a similar parallelization design, the upcoming GROMACS version will also support the Intel MIC accelerator architecture. Intel MIC supports native execution of standard MPI codes using the so-called symmetric mode, where the card is essentially treated as
a general-purpose multi-core node. However, as MIC is a highly parallel architecture requiring fine-grained parallelism, many parts of typical MPI codes will be inefficient on these processors. Hence, efficient utilization of Xeon Phi devices in molecular dynamics -- especially with typical bio-molecular simulations and strong-scaling in mind - is only possible by treating them as accelerators. Similarly to GPUs, this means a parallelization scheme based on offloading only those tasks that are suitable for wide SIMD and highly thread-parallel execution to MIC.

\subsection{Ensemble simulations}
The performance and scaling advances in GROMACS (and many other programs) have made it efficient to run simulations that simply were too large only a few years ago. However, infrastructures such as the European PRACE provide access only to problems that scale to thousands of cores. This used to be an impossible barrier for biomolecular dynamics on anything but ridiculously large systems when an implementation could only run well with hundreds of particles per core. Scaling has improved, but the number of computational units in supercomputers is growing even faster. There are now multiple machines in the world that reach roughly a million cores. Under ideal conditions, GROMACS can scale to levels where each PP rank handles 40 atoms, but there are few if any concrete biological problems that require 40 million atoms without corresponding increases in the number of samples generated. Even in the theoretical case where we could improve scaling to the point where each core only contains a single atom, the simulation system would still be almost an order of magnitude larger than the example in Fig.~\ref{glic}.

To adapt to this reality, researchers are increasingly using large ensembles of simulations, either to simply sample better, or new algorithms such as replica exchange simulation,\cite{SugOka1999} Markov state models,\cite{Schuette2012} or milestoning\cite{Faradjian2004} that analyze and exchange data between multiple simulations to improve overall sampling. In many cases, this achieves as much as two-fold superscaling, i.e., an ensemble of 100 simulations running on 10 nodes each might provide the same sampling efficiency as a single simulation running on 2000 cores. To automate this, GROMACS has been co-developed with a new framework for Parallel Adaptive Molecular Dynamics called “Copernicus.”\cite{ProLarPou2011} Given a set of input structures and sampling settings, this framework automatically starts a first batch of sampling runs, makes sure all simulations complete (with extensive support for checkpointing and restarting of failed runs), and automatically performs the adaptive step data analysis to decide what new simulations to start in a second generation. The current ensemble sampling algorithms scale to hundreds or thousands of parallel simulations (each using up to thousands of cores even for small systems). For the first time in many years, molecular dynamics might actually be able to use all the cores available on next-generation supercomputers rather than constantly being a generation or two behind.

\subsection{Multi-level load balancing}

Achieving strong scaling to a higher core count for a fixed-size problem requires careful consideration of load balance. The advantage provided by spatial DD is one of data locality and reuse, but if the distribution of computational work is not homogeneous then more care is needed. A typical membrane protein simulation is dominated by
\begin{itemize}
\item water, which is usually treated with a rigid 3-point model,
\item a lipid membrane, whose alkyl tails are modeled by particles with zero partial charge and bonds of constrained length, and
\item a protein, which is modeled with a backbone of fixed-length bonds that require a lengthy series of constraint calculations, as well as partial charge on all particles.
\end{itemize}

These problems are well known, and are addressed in the GROMACS DD scheme via automatic dynamic load balancing that distributes the spatial volumes unevenly according to the observed imbalance in compute load. This approach has limitations because it works at the level of DD domains that must map to MPI ranks, so cores within the same node or socket have unnecessary copies of the same data. We have not yet succeeded in developing a highly effective intra-rank decomposition of work to multiple cores. We hope to address this via intra-node or intra-socket task parallelism.

One advantage of the PME algorithm as implemented in GROMACS is that it is possible to shift the computational workload between the real- and reciprocal-space parts of the algorithm at will. This makes it possible to write code that can run optimally at different settings on different kinds of hardware. The performance of the compute, communication and bookkeeping parts of the overall algorithm vary greatly with the characteristics of the hardware that implements it, and with the properties of the simulation system studied. For example, shifting compute work from reciprocal to real space to make better use of an idle GPU increases the volume that must be communicated during DD, while lowering the required communication volume during the 3D FFTs. Evaluating how best to manage these compromises can only happen at runtime.

The MPMD version of PME is intended to reduce the overall communication cost on typical switched networks by minimizing the number of ranks participating in the 3D FFTs. This requires generating a mapping between PME and non-PME ranks and scheduling data transfer to and from them. However, on hardware with relatively efficient implementations of global communication, it can be advantageous to prefer the SPMD implementation because it has more regular communication patterns.\cite{AbrGre2011} The same may be true on architectures with accelerators, because the MPMD implementation makes no use of the accelerators on the PME ranks. The performance of both implementations is limited by lack of overlap of communication and computation.

Attempts to use low-latency partitioned global address space (PGAS) methods that require single-program multiple-data (SPMD) approaches are particularly challenged, because the gain from any decrease in communication latency must also overcome the overall increase in communication that accompanies the MPMD-to-SPMD transition.\cite{ReyTurHes2013} The advent of implementations of non-blocking collective (NBC) MPI routines is promising if computation can be found to overlap with the background communication. The most straightforward approach would be to revert to SPMD and hope that the increase in total communication cost is offset by the gain in available compute time, however, the available performance is still bounded by the overall cost of the global communication. Finding compute to overlap with the NBC on the MPMD PME ranks is likely to deliver better results. Permitting PME ranks to execute kernels for bonded and/or non-bonded interactions from their associated non-PME ranks is the most straightforward way to achieve this overlap. This is particularly true at the scaling limit, where the presence of bonded interactions is one of the primary problems in balancing the compute load between the non-PME ranks.

The introduction of automatic ensemble computing introduces another layer of decomposition, by which we essentially achieve MSMPMD parallelism: Multiple-simulation (ensemble), multiple-program (direct/lattice space), and multiple-data (domain decomposition).

\subsection{Managing the long-range contributions at exascale}

A promising candidate for exascale-level biomolecular simulations is the use of suitable implementations of fast-multipole methods such as ExaFMM.\cite{YokBar2012,ArnFahHol2013} At least one implementation of FMM-based molecular dynamics running on 100,000 cores has been reported,\cite{AndNorFuj2013} but so far the throughput on problems of comparable size is only equivalent to the best PME-based implementations. FMM-based algorithms can deliver linear scaling of communication and computation with both the number of MPI ranks and the number of particles. This linear scaling is expected to be an advantage when increasing the number of processing units in the exascale era. Early tests showed that the iteration times of ExaFMM doing only long-range work and GROMACS 4.6 doing only short-range work on homogeneous systems of the same size were comparable, so we hope we can deploy a working version in the future.

\subsection{Fine-grained task parallelism for exascale}
We plan to address some of the exascale-level strong-scaling problems mentioned above through the use of a more fine-grained task parallelism than what is currently possible in GROMACS. Considerable technical challenges remain to convert OpenMP-based data-parallel loop constructs into series of tasks that are coarse enough to avoid spending lots of time scheduling work, and yet fine enough to balance the overall load. Our initial plan is to experiment with the cross-platform Thread Building Blocks (TBB) library,\cite{TBB} which can coexist with OpenMP and deploy equivalent loop constructs in the early phases of development. Many alternatives exist; those that require the use of custom compilers, runtime environments, or language extensions are unattractive because that increases the number of combinations of algorithm implementations that must be maintained and tested, and compromises the high portability enjoyed by GROMACS.

One particular problem that might be alleviated with fine-grained task parallelism is reducing the cost of the communication required during the integration phase. Polymers such as protein backbones are modeled with fixed-length bonds, with at least two bonds per particle, which leads to coupled constraints that domain decomposition spreads over multiple ranks. Iterating to satisfy those constraints can be a costly part of the algorithm at high parallelism. Because the spatial regions that contain bonded interactions are distributed over many ranks, and the constraint computations cannot begin until after all the forces for their atoms have been computed, the current implementation waits for all forces on all ranks to be computed before starting the integration phase. The performance of post-integration constraint-satisfaction phase is bounded by the latency for the multiple communication stages required. This means that ranks that lack atoms with coupled bonded interactions, such as all those with only water molecules,literally have nothing to do at this stage. In an ideal implementation, such ranks could contribute very early in each iteration to complete all the tasks needed for the forces for the atoms involved in coupled bond constraints. Integration for those atoms could take place while forces for interactions between unrelated atoms are being computed, so that there is computation to do on all nodes while the communication for the constraint iteration takes place. This kind of implementation would require considerably more flexibility in the book-keeping and execution model, which is simply not present today.

\section{Handling exascale software challenges: process and infrastructure}

\subsection{Transition from C to C++98}
The major part of the GROMACS code base has been around 1-1.5 million lines of C code since version 4.0 (http://www.ohloh.net/p/gromacs). Ideally, software engineering on such moderately large multi-purpose code bases would take place within the context of effective abstractions.\cite{WilAruBro2014} For example, someone developing a new integration algorithm should not need to pay any attention to whether the parallelization is implemented by constructs from a threading library (like POSIX threads), a compiler-provided threading layer (like OpenMP), an external message-passing library (like MPI), or remote direct memory access (like SHMEM). Equally, she/he should not need to know whether the kernels that compute the forces they are using as inputs are running on any particular kind of accelerator or CPU. Implementing such abstractions generally costs some developer time, and some compute time. These are necessary evils if the software is to be able to change as new hardware, new algorithms or new implementations emerge.

Considerable progress has been made in modularizing some aspects of the code base to provide effective abstraction layers. For example, once the main MD iteration loop has begun, the programmer does not need to know whether the MPI layer is provided by an external library because the computation is taking place on multiple nodes, or the internal thread-based implementation is working to parallelize the computation on a single node. Portable abstract atomic operations have been available as a side-effect of the thread-MPI development. Integrators receive vectors of  positions, velocities and forces without needing to know the details of the kernels that computed the forces. The dozens of non-bonded kernels can make portable SIMD function calls that compile to the correct hardware operations automatically.

However, the size of the top-level function that implements the loop over time steps has remained at about 1800 code and comment lines since 4.0. It remains riddled with special-case conditions, comments, and function calls for different parallelization conditions, integration algorithms, optimization constructs, housekeeping for communication and output, and ensemble algorithms. The function that computes the forces is even worse, now that both the old and new non-bonded kernel infrastructures are supported! The code complexity is necessary for a general-purpose multi-architecture tool like GROMACS. However, needing to be aware of dozens of irrelevant possibilities is a heavy barrier to participation in the project, because it is very difficult to understand all side effects of a change.

To address this, we are in the process of a transition from C99 to C++98 for much of this high-level control code. While we remain alert to the possibility that HPC compilers will not be as effective at compiling C++98 as they are for C99, the impact on execution time of most of this code is negligible and the impact on developer time is considerable. 

Our expectation is that the use of
virtual function dispatch will eliminate much of the complexity of understanding conditional code (including switch statements over enumerations that must be updated in widely scattered parts of the code), despite a slightly slower implementation of the actual function call. After all, GROMACS has long used a custom vtable-like implementation for run-time dispatch of the non-bonded interaction kernels.
Objects managing resources via RAII exploiting compiler-generated destructor calls for doing the right thing will lead to shorter development times and fewer problems because developers have to manage fewer things. Templated container types will help alleviate the burden of manual memory allocation and deallocation. Existing C++ testing and mocking libraries will simplify the process of developing adequate testing infrastructure, and existing task-parallelism support libraries such as Intel TBB\cite{TBB} will be beneficial.

It is true that some of these objectives could be met by re-writing in more objected-oriented C, but the prospect of off-loading some tedious tasks to the compiler is attractive.

\subsection{Best practices in open-source scientific software development}

Version control is widely considered necessary for successful software development. GROMACS used CVS in its early days and now uses Git (git clone git://git.gromacs.org/gromacs.git). The ability to trace when behavior changed and find some metadata about why it might have changed is supremely valuable.

Coordinating the information about desires of users and developers, known problems, and progress with current work is an ongoing task that is difficult with a development team scattered around the world and thousands of users who rarely meet. GROMACS uses the Redmine issue-tracking system\footnote{http://redmine.gromacs.org} to discuss feature development, report and discuss bugs, and to monitor intended and actual progress towards milestones. Commits in the git repository are expected to reference Redmine issues where appropriate, which generates automatic HTML cross-references to save people time finding information.

Peer review of scientific research is the accepted gold standard of quality because of the need for specialist understanding to fully appreciate, value, criticize and improve the work. Software development on projects like GROMACS is comparably complex, and our experience has been that peer review has worked well there. Specifically, all proposed changes to GROMACS -- even from the core authors -- must go through our Gerrit code-review website\footnote{http://gerrit.gromacs.org}, and receive positive reviews from at least two other developers of suitable experience, before they can be merged. User- and developer-level documentation must be part of the same change. Requiring this review to happen before acceptance has eliminated many problems before they could be felt. It also creates social pressure for people to be active in reviewing others' code, lest they have no karma with which to get their own proposals reviewed. As features are implemented or bugs fixed, corresponding Redmine issues are automatically updated. Gerrit also provides a common venue for developers to share work in progress, either privately or publicly.

Testing is one of the least favourite activities of programmers, who would much rather continue being creative in solving new problems. The standard procedure in software engineering is to deploy “continuous integration,” where each new or proposed change is subjected to a range of automatic tests. In the GROMACS project, we use Jenkins\footnote{http://jenkins.gromacs.org} to build the project on a wide range of operating systems (MacOS, Windows, flavours of Linux), compilers (GNU, Intel, Microsoft, clang; and several versions of each), and build configurations (MPI, thread-MPI, OpenMP, different kinds of SIMD), and then automatically test the results for correctness. This immediately finds problems such as programmers using POSIX constructs that are not implemented on Windows. Most of our tests detect regressions, where a change in the code leads to an unintended change in behavior. Unfortunately, many of these tests are still structured around executing a whole MD process, which makes it difficult to track down where a problem has occurred, unless the code change is tightly focused. This motivates the discipline of proposing changes that only have one logical effect, and working towards adding module-level testing. New behaviors are expected to be integrated alongside tests of that behavior, so that we continue to build upon the test infrastructure for the future. All tests are required to pass before code changes can be merged.

Testing regularly for changes in execution speed is an unsolved problem that is particularly important for monitoring our exascale software developments. It is less suited for deployment via continuous integration, because of the quantity of computation required to test the throughput of code like GROMACS with proper load-balancing, at-scale, and on a range of hardware and input conditions. It would be good to be able to execute a weekly end-to-end test run that shows that unplanned performance regressions have not emerged, but we have not prioritized it yet. Waiting to do these tests until after feature stability is achieved in the software-development life cycle is not appropriate, because that requires extra work in identifying the point in time (ie. the git commit) where the problem was introduced, and the same work identifying the correct way to manage the situation. This is much better done while the change is fresh in developers' minds, so long as the testing procedure is reasonably automatic. Also, in the gap between commit and testing, a regression may be masked by some other improvement. More extensive human-based testing before releases should still be done; but avoiding protracted bug hunts just before releases makes for a much happier team.

Cross-platform software requires extensive configuration before it can be built. The system administrator or end user needs to be able to guide what kind of GROMACS build takes place, and the configuration system needs to verify that the compiler and machine can satisfy that request. This requires searching for ways to resolve dependencies, and disclosing to the user what is being done with what is available. It is important that compilation should not fail when configuration succeeded, because the end user is generally incapable of diagnosing what the problem was. A biochemist attempting to install GROMACS on their laptop generally does not know that scrolling back through 100 lines of output from recursive make calls is needed to find the original compilation error, and even then they will generally need to ask someone else what the problem is and how to resolve it. It is far more efficient for both users and developers to detect during configuration that compilation will fail, and to provide suggested solutions and guidance at that time. Accordingly, GROMACS uses the CMake build system (http://www.cmake.org), primarily for its cross-platform support, but makes extensive use of its high-level constructs, including sub-projects and scoped variables.

\subsection{Profiling}
Experience has shown that it is hard to optimize software, especially an HPC code, based on simple measurements of total execution speed. It is often necessary to have a more fine-grained view of the performance of individual parts of the code, details of execution on the individual compute units, as well as communication patterns. There is no value in measuring the improvement in execution time of a non-bonded kernel if the execution time of the FFTs is dominant!

Standard practice is to use a profiling/tracing tool to explore which functions or code lines consume important quantities of time, and to focus effort on those. However, if the measurement is to provide useful information, the profiler should perturb the execution time by a very small amount.  This is particularly challenging with GROMACS because in our case an MD iteration is typically in the range of a millisecond or less wall clock time around the current scaling limit, and the functions that are interesting to profile might execute only for microseconds. Overhead introduced by performance measurement that is acceptable in other kinds of applications often leads to incorrect conclusions for GROMACS. Statistical sampling from periodically interrupting the execution to observe which core is doing which task could work in principle, but (for example) Intel's VTune 3 Amplifier defaults to a 10 ms interval, which does not create confidence that use of the tool would lead to accurate observations of events whose duration is a thousand times shorter. Reducing the profiling overhead to an acceptable level while still capturing enough information to be able to easily interpret the performance measurements has proved challenging. Additionally, this often required expert knowledge, assistance of the developers of the respective performance measurement tool. This makes it exceptionally hard  to use in-depth or large-scale profiling as part of the regular GROMACS development workflow.

However, we have not been optimizing in the dark; the main mdrun simulation tool has included a built-in tracing-like functionality for many years. This functionality relies on manual instrumentation of the entire source code-base (through inlined start/stop timing functions) as well as low-overhead timing measurements based on processor cycle counters. The  great benefit is that the log output of every GROMACS simulation contains a breakdown of detailed timing measurements of the different code parts. However, this internal tracing functionality does not reach its full potential because the collected data is typically displayed and analyzed through time-averages across MPI ranks and time-steps, often hiding useful details.

To realize more of this potential, we have explored the possibility of more detailed MPI rank-based statistics, including minimum and maximum execution times across ranks as well as averages. However, this information is still less detailed than that from a classical trace and profile visualizer. We are exploring combining our internal instrumentation with a tracing library. By adding API calls to various tracing libraries to our instrumentation calls, we can provide native support for detailed trace-generation in GROMACS just by linking against a tracing library like Extrae\footnote{http://www.bsc.es/computer-sciences/extrae}. This will make it considerably easier to carry out performance analysis without the need for expert knowledge on collecting performance data while avoiding influencing the program behavior by overhead.

\section{Future directions}
GROMACS has grown from an in-house simulation code into a large international software project, which now also has highly professional developer, testing and profiling environments to match it. We believe the code is quite unique in the extent to which it interacts with the underlying hardware, and while there are many significant challenges remaining this provides a very strong base for further extreme-scale computing development. However, scientific software is rapidly becoming very dependent on deep technical computing expertise: Many amazingly smart algorithms are becoming irrelevant since they cannot be implemented efficiently on modern hardware, and the inherent complexity of this hardware makes it very difficult even for highly skilled physicists and chemists to predict what will work. It is similarly not realistic to expect every research group to afford a resident computer expert, which will likely require both research groups and computing centers to increasingly join efforts to create large open source community codes where it is realistic to fund multiple full time developers. 
In closing, the high performance and extreme-scale computing landscape is currently changing faster than it has ever done before. It is a formidable challenge for software to keep up with this pace, but the potential rewards of exascale computing are equally large.

\section*{Acknowledgments}
This work was supported by the European research Council (258980, BH), the Swedish e-Science research center, and the EU FP7 CRESTA project (287703). Computational resources were provided by the Swedish National Infrastructure for computing (grants SNIC 025/12-32 \& 2013-26/24) and the Leibniz Supercomputing Center. 

\bibliographystyle{splncs03}
\bibliography{bibliography}
\end{document}